# Second harmonic AC calorimetry technique within a diamond anvil cell

Nathan Dasenbrock-Gammon
*Department of Physics and Astronomy, University of Rochester, Rochester, NY 14627, USA.*

Raymond McBride, Gyeongjae Yoo, and Sachith Dissanayake
*Department of Mechanical Engineering, University of Rochester, Rochester, NY 14627, USA.*

Ranga Dias*
*Department of Physics and Astronomy, University of Rochester, Rochester, NY 14627, USA. and Department of Mechanical Engineering, University of Rochester, Rochester, NY 14627, USA.*
(*Corresponding Author: rdias@rochester.edu)

Tuning the energy density of matter at high pressures gives rise to exotic and often unprecedented properties, e.g., structural transitions, insulator-metal transitions, valence fluctuations, topological order, and the emergence of superconductivity. The study of specific heat has long been used to characterize these kinds of transitions, but their application to the diamond anvil cell (DAC) environment has proved challenging. Limited work has been done on the measurement of specific heat within DACs, in part due to the difficult experimental setup. To this end we have developed a novel method for the measurement of specific heat within a DAC that is independent of the DAC design and therefore readily compatible with any DACs already performing high pressure resistance measurements. As a proof-of-concept, specific heat measurements of the $MgB_2$ superconductor were performed, showing a clear anomaly at the transition temperature ($T_c$), indicative of bulk superconductivity. This technique allows for specific heat measurements at higher pressure than previously possible.

## I. INTRODUCTION

The study of pressure induced phase transitions leading to novel phases is a very active area of research. One of the most profound techniques to observe phase transitions in materials is specific heat, due to its sensitivity to microscopic energy scales. A number of exotic physical phenomena manifest themselves with different characteristic features in the specific heat of a material; superconductivity [1], latent heat of structural phase transitions or chemical reactions [2], and magnetic ordering, to name a few. Although such measurements are common under ambient conditions, their applications under pressure have remained limited. Detecting subtle phase transitions in extreme sample environments, such as at high pressure, necessitates new techniques that are robust yet also easily adoptable.

Conventionally, at ambient conditions, specific heat is measured using an adiabatic technique whereby the sample and mount are subjected to a heat pulse and the resulting temperature decay is measured [3]. By fitting the temperature decay with heat transfer models the specific heat values can be extracted. This method works well when the samples are relatively large, and the sample mount is reasonably small. The requirements to achieve adiabatic conditions fail when higher pressures are desired. Large mechanical assemblies, such as DACs or clamp cells, are required to apply high pressures, and necessitate smaller sample sizes. In DAC experiments, at the highest pressures, the sample mass will be many orders of magnitude smaller than the DAC mass. Therefore, different techniques are required for specific heat measurements under pressure.

There are several studies on thermal measurements of materials under pressure within clamp cells [4, 5], however, these apparatus have a limited amount of pressure that they can generate. While clamp cells are limited to pressures of only a few GPa, DACs are capable of generating pressures well into the megabar regime [6]. The ability to perform specific heat measurements within a DAC allows measurement of thermal properties at higher pressures than clamp cells can obtain, yet, to date, measurements of specific heat within DACs have remained limited.

In the most basic form, specific heat measurements are performed through the application of heat and measurement of the resulting change in temperature. A few methods have been used to perform such measurements inside a DAC. One such method is the chopped laser heating technique whereby a laser beam is used to pulse heat a sample within the DAC while the temperature fluctuations are measured [7]. Another method is the high frequency third harmonic technique [8], whereby a single heater wire is used to both heat the sample and extract thermal properties from the third harmonic response to measure specific heat within a DAC [9, 10]. This method has been successfully employed for measurements of the specific heat of water up to 9 GPa within a DAC using a thin piece of gold foil as both heater and thermometer [11]. In contrast to this third harmonic method, we will present a second harmonic technique whereby a Joule heater connected to the sample is driven at frequency $f$, resulting in small oscillations of the sample temperature at $2\times f$ that can be measured by a thermocouple attached to the sample. If the temperature oscillations are small, the specific heat can thus be measured at any externally maintained temperature. Such second harmonic techniques are common at ambient pressures [12], and has experienced some use in the aforementioned clamp cells [4, 5]. A similar technique has been employed in Bridgman-type cells [13], and the author speculated about the applicability within a DAC.

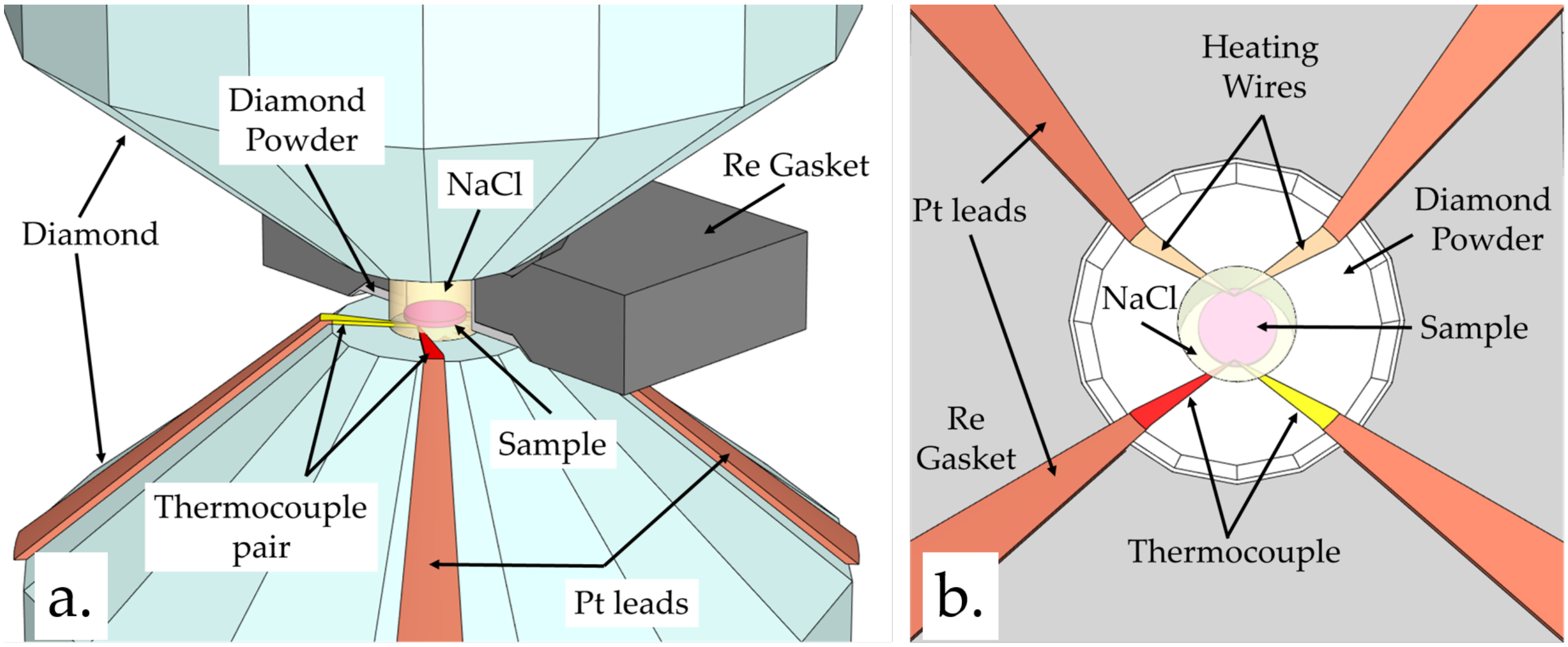

FIG. 1. Schematic rendering of the novel AC calorimetry technique (not to scale). The sample is surrounded by an NaCl insert with a heater and thermocouple making contact with the sample. **a.** View of the preparation as seen from the side showing the thermocouple making contact with the sample inside the DAC. **b.** View of the preparation as seen from the top of the sample area showing the configuration of heater, thermocouple, and Pt leads.

In this report we present a novel technique of measuring specific heat within a DAC. This technique relies on modifications of standard conductivity DAC techniques [14, 15], allowing for easy adaptation and use. In this method, a Joule heater and thermocouple are connected directly to the sample. By measuring the second harmonic voltage response on the thermocouple with respect to the heater drive, a semi-quantitative specific heat value is calculated. This method is demonstrated by carrying out measurements around the superconducting transition of $MgB_2$ at 1.5 GPa.

## II. EXPERIMENTAL METHODS

### A. DAC preparation

The modified AC calorimetry technique involves incorporating a thermocouple and heater in contact with a sample directly inside a DAC under pressure. Figure 1 shows schematic views of the DAC preparation used for this technique. As this preparation can be considered an adaptation of standard conductivity techniques, the preparation is similar. First, a gasket is preindented, drilled, and remounted onto one of the diamonds. The gasket is then packed with diamond or alumina powder ($Al_2O_3$) to form an electrically insulating layer. Next, a small chamber is carved out and filled with an NaCl insert. The NaCl functions as both a pressure transmitting medium and thermal/electrical insulation in which the sample will be placed. Two sets of shorted contacts are placed into the NaCl insert. These contacts are cut out of thin foils into triangular shapes. The first consists of a thermocouple pair, such as alumel/chromel, though other thermocouple pairs of sufficiently differing Seebeck coefficients could also be used. The second set of contacts is the resistive heater, here Ti was used. Figure 2 shows a setup with the thermocouple pair and heater prepared (**a.**) The sample is then placed in between the two sets of shorted contacts, as shown in figure 2 (**b.**), and upon DAC compression will form pressure contact with the sample. An additional piece of NaCl is placed on top of the sample to thermally isolate it from the diamond. A ruby piece can be placed near the sample for pressure measurement using ruby fluorescence [16, 17]. To finish the setup, Pt leads are connected to the back end of the heater and thermocouple to run out of the sample chamber up the facets, where they are soldered to larger copper wires allowing for connection to electronic equipment.

### B. Principle of operation

The governing equation for AC calorimetry is given by [12],

$$C_{AC} = \frac{P}{\omega T_{AC}} F(\omega), \tag{1}$$

where $C_{AC}$ is the AC specific heat, $P$ the driving power, $\omega$ the driving frequency, $T_{AC}$ the small temperature change experienced by the sample due to the heater, and $F(\omega)$ the frequency response curve. The frequency response curve is given by

$$F(\omega) = \frac{1}{\sqrt{1+(\omega_1/\omega)^2+(\omega/\omega_2)^2}}. \tag{2}$$

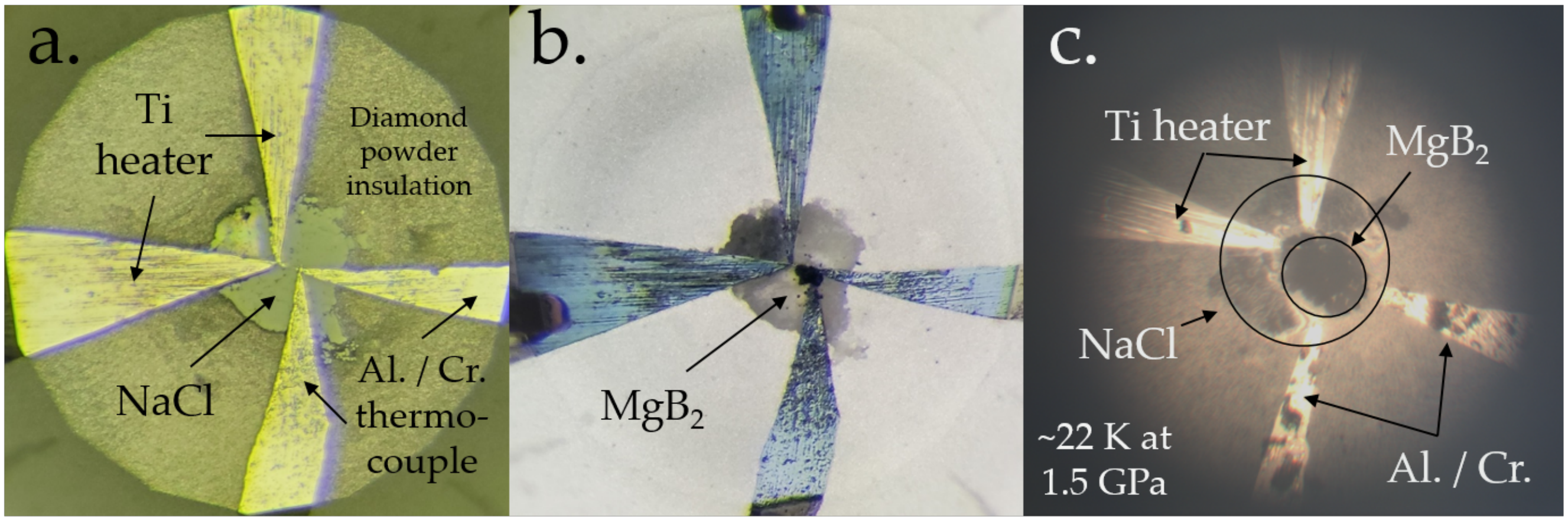

FIG. 2. Microimages of the AC calorimetry setup. **a.** A DAC setup before sample loading. A preindent is packed with insulating diamond powder and a center sample area filled with an NaCl insert to act as thermal insulation and pressure transmitting medium. Two titanium strips are shorted to form the heater, while an alumel / chromel pair are shorted to form a thermocouple. **b.** The $MgB_2$ sample is placed in between the heater and thermocouple. **c.** An image of a different preparation taken at 1.5 GPa and 22 K showing all components of the AC calorimetry setup.

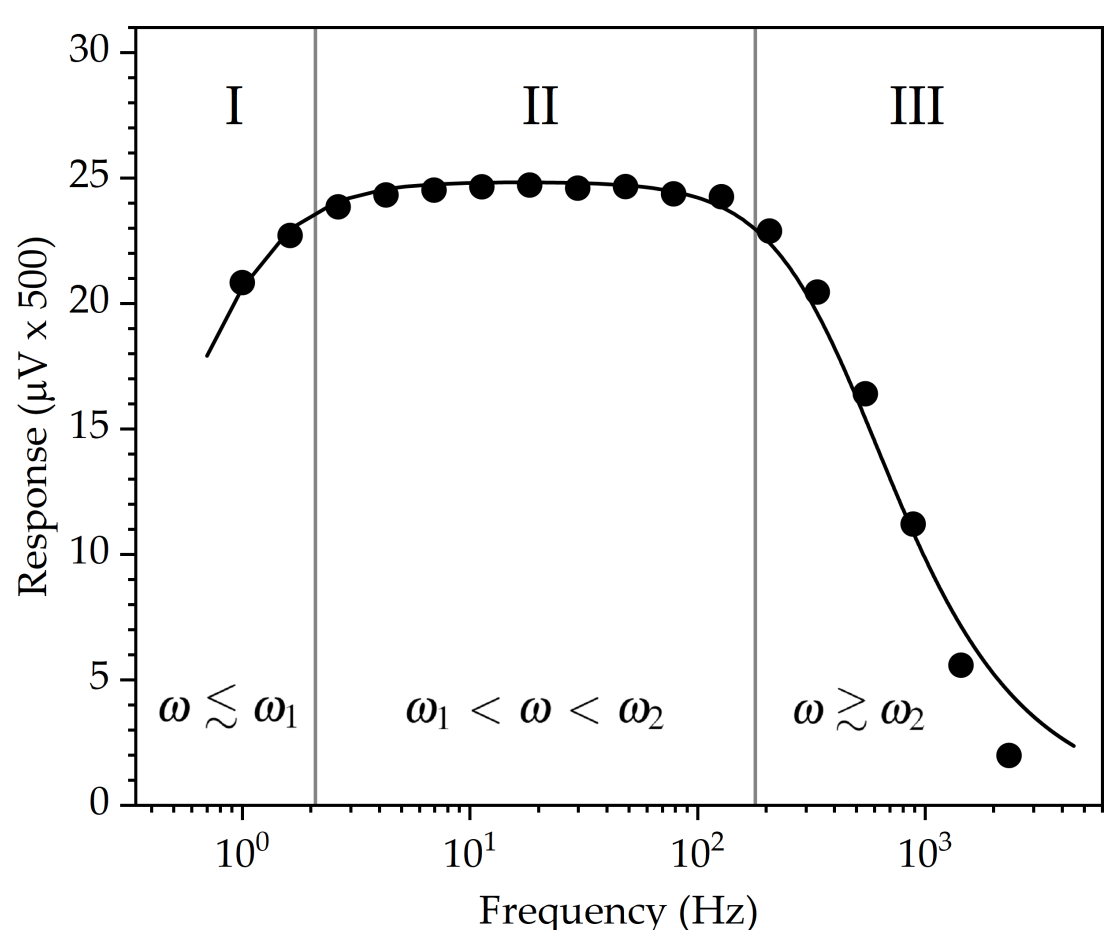

FIG. 3. Plot of a frequency sweep performed on a sample at room temperature that illustrates the three regimes of $F(\omega)$ as the driving frequency is changed relative to $\omega_1$ and $\omega_2$. The plotted curve is a fit to the frequency response equation multiplied by a magnitude prefactor.

$F(\omega)$ depends on the relative timescales of the system: $\omega$, $\omega_1$, and $\omega_2$, the driving frequency, frequency of thermalization between the sample and the environment, and frequency of self thermalization of the heater-sample-thermocouple assembly, respectively. $F(\omega)$ can be divided into three regimes depending on the relative value of $\omega$ as compared with $\omega_1$ and $\omega_2$, as described in [18]:

I. $\omega \lesssim \omega_1$, the driving frequency is slower than or equivalent to the timescale of thermal equilibration with the environment. Heat is dissipated into the environment before the sample is able to thermally respond, thus $F(\omega) < 1$, and as the frequency increases there is improved response until $\omega \approx \omega_1$.

II. $\omega_1 < \omega < \omega_2$, the driving frequency is intermediate to the timescales of equilibration with the environment and thermalization of the heater-sample-thermocouple assembly. There is optimal thermal coupling in the system and $F(\omega) \approx 1$. The response is relatively flat with respect to the driving frequency in this regime.

III. $\omega \gtrsim \omega_2$, the driving frequency is faster than or equivalent to the timescale of thermalization of the heater-sample-thermocouple assembly. The heater is changing in temperature faster than the sample is able to thermally respond. $F(\omega) < 1$ and decreases as the driving frequency is increased.

While $\omega$ is determined by the experimenter in choosing the driving frequency, $\omega_1$ and $\omega_2$ depend on the precise geometry of a preparation and will vary slightly between experiments. However, it is essential that the heater-sample-thermocouple assembly be able to thermalize faster than heat is able to leak into the environment; that is, $\omega_1 << \omega_2$. To this end, the sample is surrounded with NaCl to better thermally isolate the heater-sample-thermocouple assembly. Due to $\omega_1$ and $\omega_2$ differing slightly between preparations, a frequency sweep must be performed to determine the optimal driving frequency. Figure 3 shows the frequency behaviour of a sample taken at room temperature along with a fitted curve to the frequency response function (multiplied by the magnitude prefactor) that clearly illustrates the three regimes.

In these experiments it is difficult to determine $P$ and $T_{AC}$, as both will depend on many factors and the various couplings and specifics of each setup, as well as likely be both pressure and temperature dependent. Because of this, it suffices to consider $C_{AC}$ in terms of arbitrary units instead of direct quantitative measurements. As we will show later with our test sample $MgB_2$, this limitation does not restrict the power of this technique in observing features in the specific heat for the identification of phase transitions.

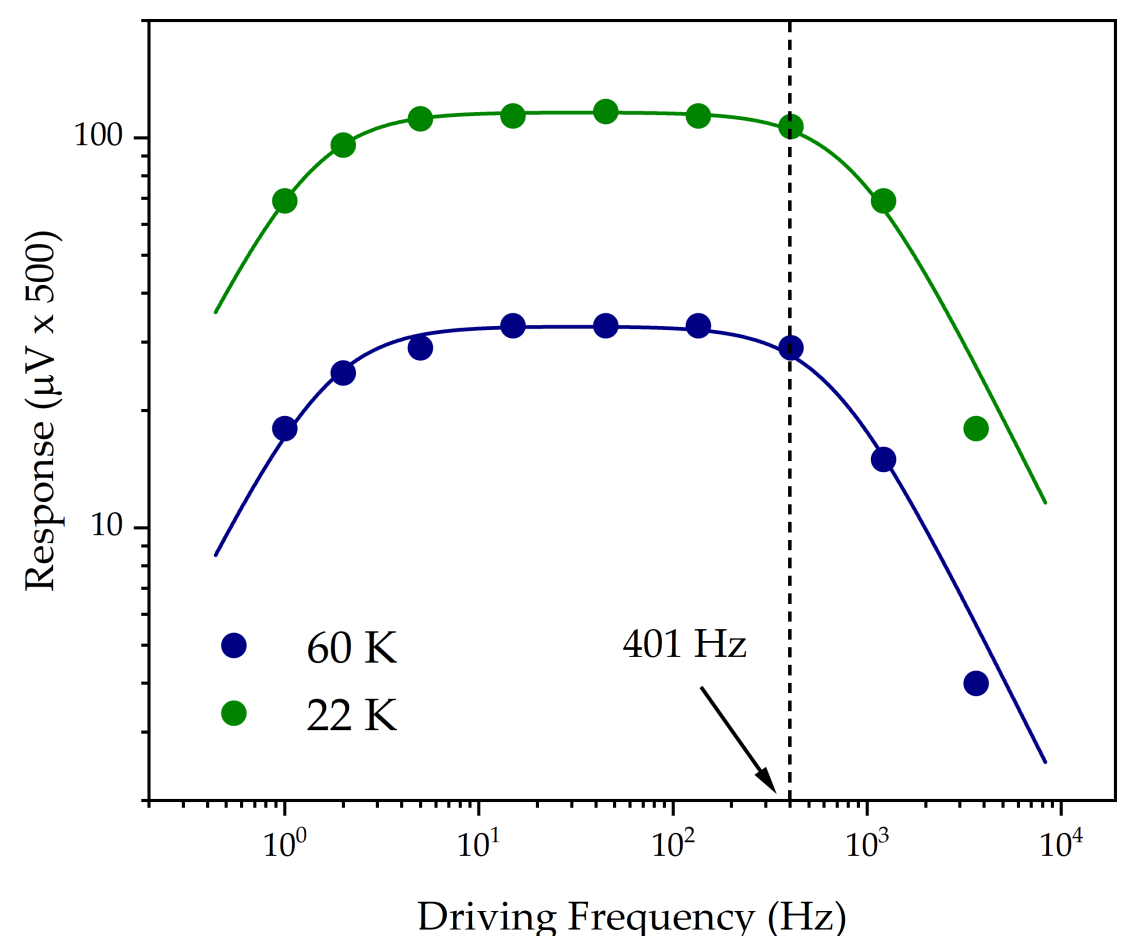

FIG. 4. Frequency sweeps taken at 60 K and 22 K on the $MgB_2$ above and below the transition temperature used to determine the proper frequency for measurement. The curves are fits to the frequency response function multiplied by a magnitude prefactor.

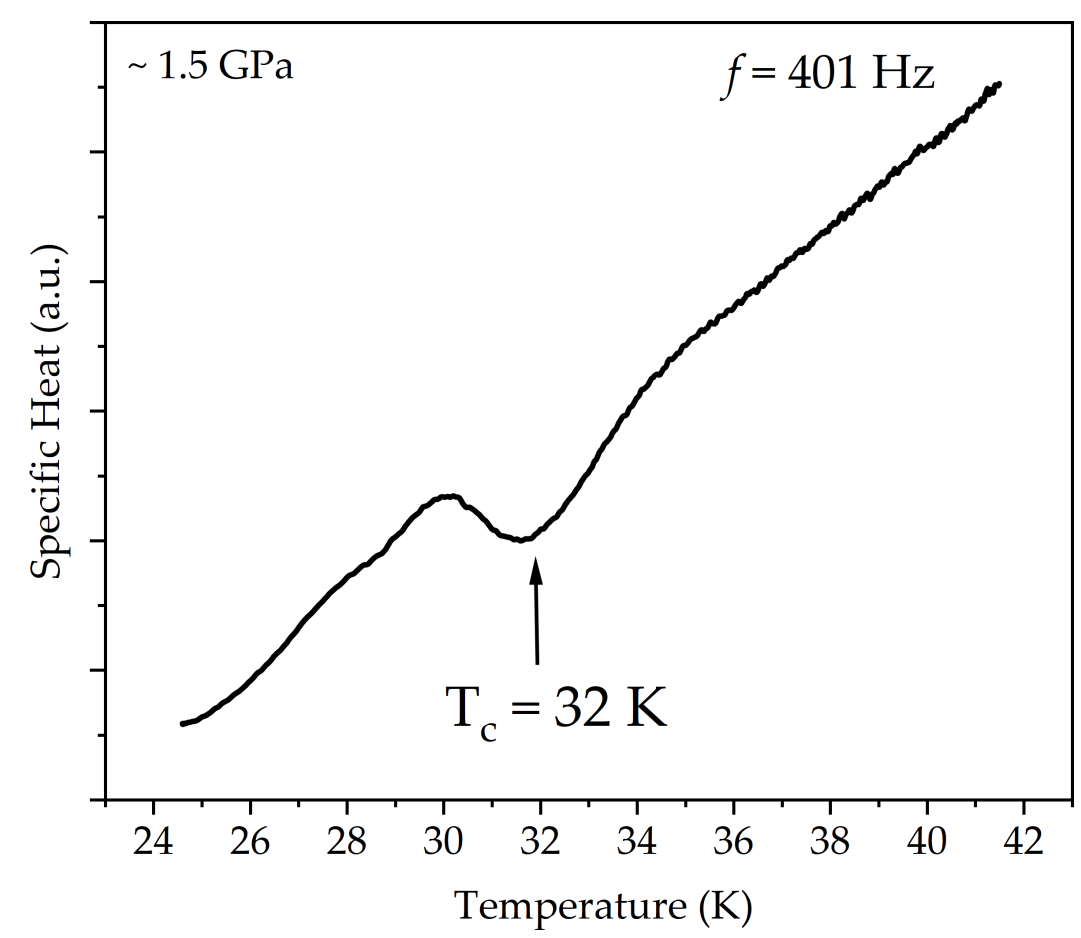

FIG. 5. Specific heat measurement performed on $MgB_2$ at 1.5 GPa. The heater was driven with 10 mA at 401 Hz. A clear rise in specific heat is seen at 32 K, corresponding to the superconducting transition.

Since the heater is driven by Joule heating, $P \propto I_\omega^2$, when driven with current $I_\omega$ , the temperature variation of the sample $T_{AC}$ will oscillate with frequency $2 \times \omega$. This results in voltage oscillations across the thermocouple at $2 \times \omega$. Recalling eq. 1, with $T_{AC} \propto V_{2\omega}$ , where $V_{2\omega}$ is the second harmonic voltage as measured by the lock in amplifier, we get:

$$C_{AC} \propto \frac{1}{V_{2\omega}}. \quad (3)$$

We are thus able to semi-quantitatively measure the specific heat of our sample within a DAC by examining the inverse of the voltage measured across the thermocouple at the second harmonic of the applied current across the heater.

## III. PROOF-OF-CONCEPT: SUPERCONDUCTING TRANSITION OF $MgB_2$

To demonstrate the power of this technique, a proof-of-concept experiment was performed on the superconductor $MgB_2$. Discovered in 2001 [19], $MgB_2$ is well explained by BCS theory, and as such an abrupt change in specific heat occurs at the transition temperature [20], making it a prime candidate to test the validity of the technique.

We prepared a DAC using the technique described in section II with an $MgB_2$ superconducting sample using 800 micron diameter diamonds with a rhenium gasket, see Figure 2 (**c.**). The heater was cut out of 5 micron thick Ti foil, and the thermocouple from pieces of Alumel and Chromel wire that had been hand rolled to ~20-30 micron thick using a machine wheel roller. The heater was driven using a Keithley 6221 current source, and the thermocouple response measured using an SR860 lockin amplifier with a 500x SR554 preamplifier.

As described in the previous section and illustrated in Figure 3, it is crucial to choose the appropriate frequency for each experiment. To this end we performed frequency sweeps on the $MgB_2$ sample at both 22 K and 60 K, below and above the transition temperature, respectively. In choosing a frequency we seek to maximize the response which leads to a frequency in region II. Furthermore, the use of higher frequencies allows for improved signal to noise ratios. To this end, we used a frequency on the border between regions II and III. As a final consideration, a frequency should be chosen so as to avoid any harmonic response to interference from the lab environment. In Figure 4 the two frequency sweeps are shown, and we determined 401 Hz to be the ideal frequency for this experiment.

With the frequency chosen, we performed a specific heat measurement on $MgB_2$ during the natural warming cycle at ~ 1.5 GPa. A small offset in the transition temperature was observed using 10 mA driving current which can be attributed to local DC heating. The temperature was corrected from a repeated measurement using a lower, 4 mA, driving current. As seen in Figure 5, there is a distinct feature in the specific heat at 32 K, indicative of the superconducting transition. This abrupt rise in specific heat corresponds to the formation of Cooper Pairs as the $MgB_2$ undergoes a superconducting transition [1, 20]. Through the measurement of this feature we have thus demonstrated the ability of this technique to measure the specific heat within a DAC under pressure and its unique ability to study phase transitions.

## IV. DISCUSSION AND CONCLUSIONS

In light of recent progress on high temperature hydride superconductors under pressure [14, 15, 21, 22], it is the authors' hope that this technique may provide additional characterization of their superconducting properties. As stated, this technique shares many commonalities with standard four-probe DAC conductivity techniques, with the adaptation of constructing the contacts in shorted pairs of a heater and thermocouple instead of the typical 4 probe Van der Pauw geometry, and using a second harmonic AC technique as opposed to resistance measurements. Therefore, this technique should be compatible with most DAC designs, and accessible to any labs already performing resistive measurements without the need for additional equipment. In section III we demonstrated the technique at 1.5 GPa using 800 micron diameter diamonds, though pressures far higher could be achieved using smaller diamond culets. In addition to the measurement of superconducting transitions as demonstrated on $MgB_2$, this technique should be capable of measuring phase transitions with features in the specific heat, such as structural phase transitions, melting behaviour, magnetic ordering, etc.

We have presented a novel DAC technique allowing for the measurement of specific heat within a DAC under pressure and demonstrated its ability to measure the superconducting transition of $MgB_2$ at $\sim$1.5 GPa. This technique should allow for the measurement of thermal properties under higher pressures than previously available by bringing AC calorimetric techniques into DACs using a widely accessible and adaptable technique.

### ACKNOWLEDGMENTS

We thank Anushika Athauda for useful scientific discussions. This work was supported by National Science Foundation, Grant No. DMR-1809649.

### DATA AVAILABILITY STATEMENT

The data that support the findings of this study are available from the corresponding author upon reasonable request.